# Elementary specific spin and orbital moments of ultrathin CoFeB amorphous films on GaAs(100)

Yu Yan[1,2], Cong Lu[2], Hongqing Tu[3], Xianyang Lu[4], Wenqing Liu[1,2], Junlin Wang[2], Iain Will[2], Balati Kuerbanjiang[4], Vlado K. Lazarov[4], Jing Wu[1,4], Johnny Wong[1], Biao You[3], Jun Du[3*], Rong Zhang[1] and Yongbing Xu[1,2*]

[1.] *York-Nanjing International Center of Spintronics (YNICS), Collaborative Innovation Center of Advanced Microstructures, School of Electronic Science and Engineering, Nanjing University, Nanjing 210093, China*

[2.] *Spintronics and Nanodevice laboratory, Department of Electronics, The University of York, YO10 3DD, United Kingdom*

[3.] *School of Physics, Nanjing University, Nanjing 210093, China*

[4.] *Department of Physics, The University of York, YO10 3DD, United Kingdom*

## Abstract

Nanoscale CoFeB amorphous films have been synthesized on GaAs(100) and studied with X-ray magnetic circular dichroism (XMCD) and transmission electron microscopy (TEM). We have found that the ratios of the orbital to spin magnetic moments of both the Co and Fe in the ultrathin amorphous film have been enhanced by more than 300% compared with those of the bulk crystalline Co and Fe, and in specifically, a large orbital moment of 0.56 $\mu_B$ from the Co atoms has been observed and at the same time the spin moment of the Co atoms remains comparable to that of the bulk hcp Co. The results indicate that the large uniaxial magnetic anisotropy (UMA) observed in the ultrathin CoFeB film on GaAs(100) is related to the enhanced spin-orbit coupling of the Co atoms in the CoFeB. This work offers experimental evidences of the correlation between the UMA and the elementary specific spin and orbital moments in the CoFeB amorphous film on the GaAs(100) substrate, which is significant for spintronics applications.

*) Authors to whom correspondence should be addressed.  Electronic mails: jdu@nju.edu.cn and yongbing.xu@york.ac.uk

The magnetic amorphous CoFeB alloys have attracted renewed interests for the applications in the next generation spintronics such as magnetic random access memory (MRAM)[1,2,3] and spin field effect transistor (SpinFET)[4,5]. For the development of SpinFET, the structure and magnetic properties of various ferromagnetic (FM) thin films on top of semiconductors (SC) such as GaAs and Si have been extensively studied over the last two decades[6,7,8,9,10]. One of the most interesting discoveries is a uniaxial magnetic anisotropy (UMA) observed in several FM/SC [11,12] when the thickness of the FM layer is reduced down to nanometer scale. For example, the bcc Fe films on GaAs(100) substrates display the UMA from 1.4 nm to 11.5 nm[13], and for bcc CoFe on GaAs(100), the UMA has been found between 1.1 and 1.7nm[14]. In the crystalline FM/SC systems, the magnetocrystalline anisotropy (MCA) might also change with the reduction of the thickness.[15] Generally, the UMA and MCA have been found to co-exist in most of the common FM/SC film systems[16]. To exclude the contribution from MCA, and thus focus on the UMA in the FM/SC film system, an effective method would to be to alloy metalloid material into the ferromagnetic films and to have amorphous magnetic thin films. Approximately 20% Boron alloyed with CoFe compound has been proven desirable. The additional Boron only slightly reduces the Curie temperature and saturation field while completely destroy its crystallinity [17]. Recent research indeed found that the amorphous CoFeB films deposited on top of GaAs still exhibit the UMA[18, **Error! Bookmark not defined.**]. Several models have been proposed including, bond-orientational anisotropy (BOA)[6,19], Neel-Taniguchi directional pair-ordering model[20] and random anisotropy model[21], to explain the origin of the UMA in CoFeB/GaAs. According to the BOA model, a medium-to-long range microstructural anisotropy is responsible for the UMA. The Neel-Taniguchi directional pair-ordering model introduces anisotropy via the dipole-like coupling between individual atom-pairs, leading to anisotropic chemical ordering of near-neighbour atoms in randomly oriented coordination. The random anisotropy model emphasizes the break of the rotational symmetry of the Hamiltonian, which gives rise to the hard magnetic behaviour even in random amorphous magnets. The origin of the UMA has also been suggested as being due to the enhanced spin-orbit coupling and interface interaction[22], which is controlled by the orbital moment and the crystal lattice[23]. The orbital moment has been found to have a more important role than the spin moment in giving rise to the magnetic anisotropy[28,24]. Hindmarch et al[17] compared the UMA of CoFeB on different substrates of AlGaAs/GaAs and AlGaAs. As included in table I, they found a much stronger UMA (50 Oe) on AlGaAs/ GaAs substrate associated with an enhancement of the orbital to spin magnetic moments ratios $m_{ratio}$ of both the Fe and Co sites, and the uniaxial magnetic anisotropy field $H_k$ of the UMA was found to be proportional to the $m_{ratio}$. Very recently, we have found a larger $H_k$ up to 270 Oe in a CoFeB/GaAs(100) system, which is much stronger than any previously reported values [25]. In this letter, we report a study of the spin and orbital moments in this CoFeB/GaAs(100) system using XMCD along with TEM and VSM. The large UMA observed in the CoFeB/GaAs(100) calls for a closer study of the spin and orbital moments and the spin-orbital coupling, which may play an important role in this system. It is well known that orbital angular moment plays a dominant role in determining the strength of magnetocrystalline anisotropy[Error! Bookmark not defined.] and XMCD technique is capable of probing directly the elementary specific orbital and spin moments[26, 27, 28].

The $Co_{56}Fe_{24}B_{20}$ films were grown atop GaAs substrates. The GaAs substrate orientation is (100) plane and the major flat is along the [110] direction. Before deposition of the CoFeB film, the substrate surface was etched and cleaned. First the contaminants of substrate surface were removed using acetone, ethanol and deionized water. The second step was to remove the oxide layer by immersion of the substrate into an $HCl/H_2O$ (1:1) solution for 50 s. The third step is surface reconstruction to create a flat surface for film deposition. The cleaned substrate was loaded into an ultrahigh-vacuum chamber with a base pressure lower than $6\times10^{-7}$ mbar and heated to 450 °C for 15 min and a further 30 min at 580 °C (annealing pressure lower than $8\times10^{-8}$ mbar) to obtain a clean and smooth surface[29]. The surface is allowed to cool to room temperature prior to film growth. The CoFeB films were prepared by DC magnetron sputtering deposition in 0.3 Pa argon (99.99%) at room temperature with a base pressure lower than $8\times10^{-6}$ Pa. A target containing $Co_{56}Fe_{24}B_{20}$ was used to deposit the magnetic CoFeB layer, with a thickness of 3.5 nm. Then a 2 nm Ta film was deposited as a capping layer to prevent the CoFeB film from oxidization.

Structural properties of the grown films were studied by JEOL 2200FS double aberration corrected (scanning) transmission electron microscope (S) TEM. Cross-sectional TEM specimens were prepared using conventional methods that include mechanical thinning and polishing followed by Ar ion milling in order to achieve electron transparency[37].

The in-plane magnetic hysteresis ($M$-$H$) loops were measured using a superconducting quantum interference device-vibrating sample magnetometer (SQUID-VSM). As a strong uniaxial anisotropy field ($H_k$) as large as 270 $Oe$ was expected, the VSM measurement was conducted using a maximum magnetic field of 400 $Oe$ to ensure the samples were fully saturated. The samples were measured at angles 0° and 90°, *i.e.* along the hard and easy axis, respectively.

XMCD measurements were performed at normal incidence to the Ta/CoFeB/GaAs(100) sample in the MAX Lab I1011 station. The XMCD spectra were measured at both positive and negative applied fields[30]. The data was collected by Total Electron Yield (TEY) detector in the analysis chamber under a magnetic field of 2000 $Oe$ . This was the operational limit of the magnet in the station, as the magnetic field has to be set at a relatively low value in order to limit magnet overheating[31]. The magnetization hysteresis loop of the CoFeB film along the out-of-plane direction was measured by a Polar MOKE, and the saturation field was found be to as large as 12000 $Oe$. It is apparent, from Fig.1(b), that for the out-of-plane direction, the magnetic field used during the XMCD measurements was not sufficient to saturate the sample. It is for this reason that, the spin and orbital moments obtained from the XMCD were scaled up. During this work, all the measurements were performed at room temperature.

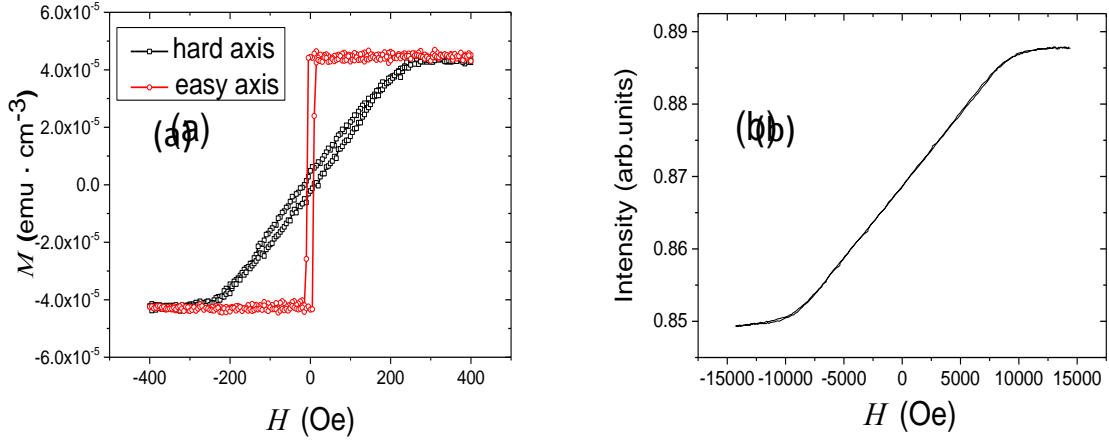

Fig.1. (a) In-plane $M$-$H$ loops along both the EA (easy axis) and HA (hard axis) for the CoFeB film deposited on GaAs(100) substrates by VSM measurement. $M$ and $H$ represent the magnetic moment and applied magnetic field, respectively. Figure (b) shows the Polar MOKE loop for the out-of-plane direction of the CoFeB film.

Fig.1(a) shows the in-plane magnetic hysteresis ($M$-$H$) loop measured by VSM along the EA and HA for the CoFeB/GaAs(100) sample. The figure indicates a clear UMA with a well-defined EA and HA axis. The value of the UMA field ($H_k$) can be obtained from the saturation field along the HA direction. Furthermore, the effective uniaxial anisotropy constant $K_u^{eff}$ can be calculated by[17]

$$K_u^{eff} = (H_k \cdot M_s)/2 \qquad (1)$$

where the $M_s$ is saturation magnetization and $H_k$ is the saturation field along the HA.

It can be seen from the Fig.1 (a) that $H_k$ has a value of $270\ Oe$, confirming our previous observation of a large UMA in the CoFeB/GaAs(100) system. According to the saturation moment and thickness measurement, the value of $M_s$ is estimated to be 976.47 Gs. The value of $K_u^{eff}$ is thus determined to be 13 $/m^3$, which is much larger than the reported values of $K_u^{eff}$=2 $kJ/m^3$ and 8 $kJ/m^3$ by Ref [17] and [35] respectively. We would also like to note that in comparing the $K_u^{eff}$ value of the CoFeB films on other GaAs orientations, *e.g.* GaAs(110) and GaAs(111) substrates[22,22,32], the $K_u^{eff}$ of the CoFeB on the GaAs(100) orientation has obvious larger value.

The hysteresis loop for the CoFeB/GaAs(100) sample along the perpendicular direction measured by a Polar MOKE is included in Fig.1(b), which shows that the perpendicular direction is the hard axis. As mentioned earlier, when making the XMCD measurement along perpendicular direction, the applied magnetic field of 2000 $Oe$ was not large enough to saturate the sample. From the perpendicular loop in figure 1(b), the saturation magnetic field is determined to be 10189 $Oe$. Comparing the magnetization at 2000 Oe and that at saturation, the data of spin and orbital moments from the XMCD have been scaled up by a factor of 5.09 as included in table I.

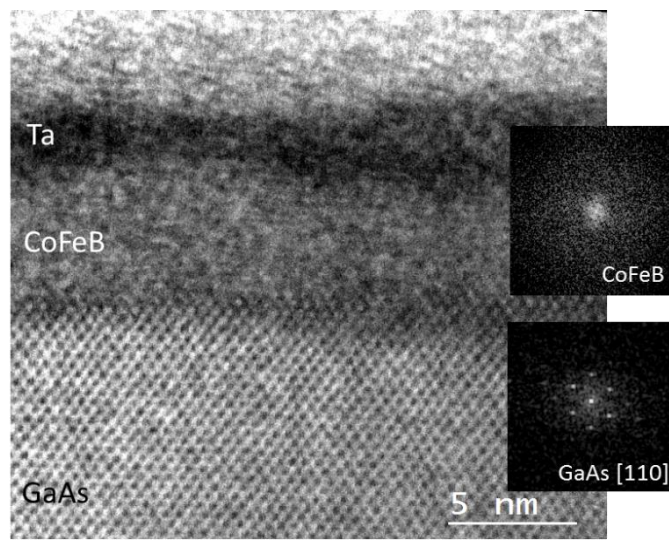

Fig.2. Cross-sectional bright-field scanning TEM micrograph of CoFeB/GaAs(100) in [110] view. The amorphous nature of the CoFeB is clearly shown by the inset digital diffractogram calculated from film area in contrast to single crystal structure of the GaAs shown by atomic planes cross fringes and Bragg reflections in the digital diffractogram (inset).

High resolution cross-sectional TEM image of the structure is shown in Fig.2. The films thicknesses of 3.5 nm and 2 nm for CoFeB and Ta, respectfully, match the growth settings. The structure of CoFeB film is amorphous starting from the very interface, in contrast to Ref [33] where interfacial crystallinity at the SC/FM interface could not be ruled out. The clear distinction between the Ta and CoFeB can be observed as well as between the CoFeB and GaAs due to single crystal structure of the GaAs substrate and the amorphous state of the CoFeB. Structural TEM analysis suggests that the interface interaction and shape anisotropy related to any possible deformation of the CoFeB layer would play little role on the formation of the UMA.

X-ray absorption spectra (XAS) of the Co and Fe $L_2$ and $L_3$ edges for CoFeB on GaAs(100) are shown in Fig. 3 (a) and (c) respectively, in which $u_+$ and $u_-$ are the absorption coefficients under antiparallel and parallel magnetic fields to the photon incident direction. Figure 3 shows the XMCD spectra for the Fe and Co $L$ - edges of the CoFeB film. According to XMCD sum rules, the orbital ($m_{orb}$) and spin ($m_{spin}$) magnetic moments and the ratio ($m_{ratio}$) of $m_{orb}$ to $m_{spin}$ can be determined from XAS and XMCD spectra by the following equations[30]:

$$m_{orb} = -\frac{4\int_{L3+L2}(u_+-u_-)d\omega}{3\int_{L3+L2}(u_++u_-)d\omega}(10-n_{3d})(1+\frac{7\langle T_Z\rangle}{2\langle S_Z\rangle}) \quad (2)$$

$$m_{spin} = -\frac{6\int_{L3}(u_+-u_-)d\omega - 4\int_{L3+L2}(u_+-u_-)d\omega}{\int_{L3+L2}(u_++u_-)d\omega} \times (10-n_{3d})(1+\frac{7\langle T_Z\rangle}{2\langle S_Z\rangle}) \quad (3)$$

$$m_{ratio} = \frac{m_{orb}}{m_{spin}} \quad (4)$$

Where $m_{orb}$ and $m_{spin}$ are the orbital and spin magnetic moments in units of $u_B/atom$, respectively, and $n_{3d}$ is the 3d electron occupation number of the specific transition metal atom. $L_3$ and $L_2$ denote the integration ranges. $\langle T_Z\rangle$ is the expectation value of magnetic dipole and $S_z$ is equal to half of $m_{spin}$ in Hartree atomic units[22]. The spin and orbital moments are also dependent on the d-band hole density in CoFeB and the intensity of the polarized x-ray in XMCD measurement. The value of $n_{3d}$ for Fe and Co in CoFeB sample is controversial. While in Ref [34], the $n_{3d}$ in amorphous film is unknown, Ref [35] gives the values of $n_{3d}$ for Fe and Co of 6.61 and 7.51, respectively, by first-principles calculation, which is the same as the reported values for bulk Fe and Co[30]. In this work, we have used the values of $n_{3d}$ from Ref [34] to calculate the spin and orbital moments of the Co and Fe in the CoFeB film.

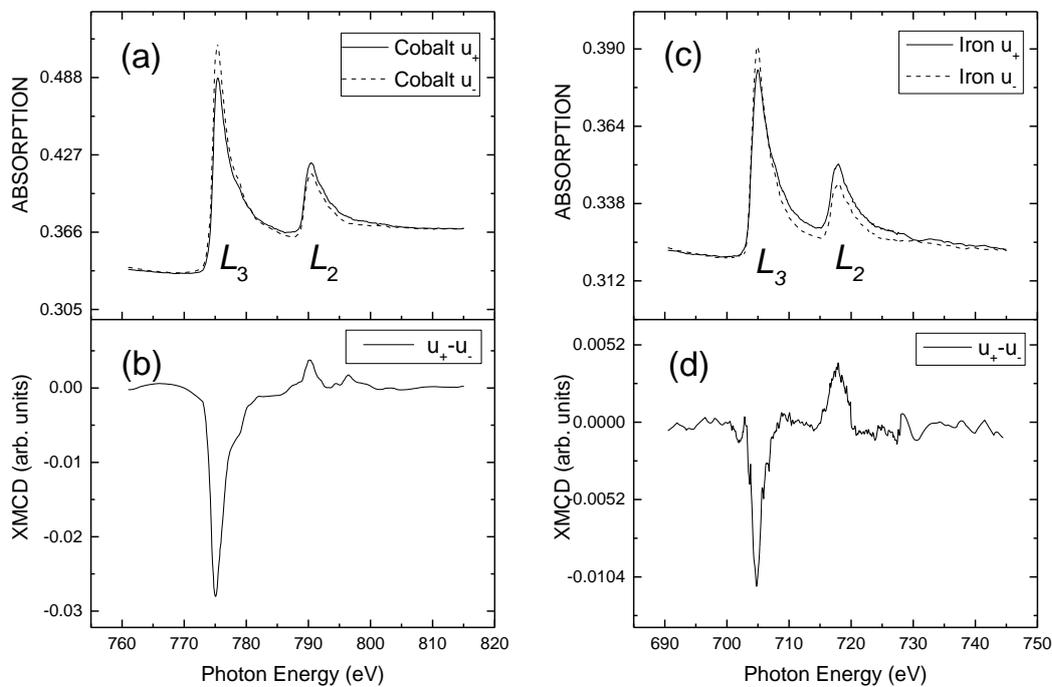

Fig.3. XAS and XMCD spectra of the Co and Fe atoms at the $L_2$ and $L_3$ edges in the CoFeB/GaAs(100 ): (a) and (c) are the XAS absorption spectra and (b) and (d) are the XMCD for Co and Fe, respectively.

The values of $m_{orb}$, $m_{spin}$ and $m_{ratio}$ are all determined from XMCD data, and the results are listed in Table I along with those reported in the literatures. Firstly, from comparing the amorphous films and the crystalline elements, one result is confirmed: the orbital moments of the Fe and Co in the amorphous films are larger than that of the crystalline bcc Fe and hcp Co. While the spin moment of the Fe atoms in the CoFeB is much reduced as compared with that of bcc Fe, the spin moment of the Co atoms remains as large as 1.53 $\mu_B$ almost the same as that of the hcp Co. As shown in table I, the orbital to spin ratios $m_{ratio}$ of the Co and Fe in the amorphous CoFeB film have been enhanced by 300% as compared with those of the hcp Co and the bcc Fe.

TABLE I. Orbital moments, spin moments and orbit to spin ratio of the Fe and Co from various CoFeB samples in units of $\mu_B/atom$.

| Sample | UMA(Oe) | Element | $m_{orb}(\mu_B)$ | $m_{spin}(\mu_B)$ | $m_{ratio}$ |
|---|---|---|---|---|---|
| Ta/Co$_{56}$Fe$_{24}$B$_{20}$(3.5nm)/GaAs(100) | 270 | Fe | 0.30 ± 0.03 | 1.17 ± 0.03 | 0.26 |
|  |  | Co | 0.56 ± 0.03 | 1.53 ± 0.03 | 0.36 |
| Ta/Co$_{40}$Fe$_{40}$B$_{20}$(3.5nm)/GaAs(100)/AlGaAs(100) (Ref.14) | 50 | Fe |  |  | 0.45 |
|  |  | Co |  |  | 0.38 |
| Ta/Co$_{40}$Fe$_{40}$B$_{20}$(3.5nm)/AlGaAs(100) (Ref.14) | 25 | Fe |  |  | 0.34 |
|  |  | Co |  |  | 0.19 |
| Ta/Co$_{40}$Fe$_{40}$B$_{20}$(2.0nm)/MgO (Ref.31) |  | Fe | 0.27 ± 0.03 | 1.77 ± 0.03 | 0.15 |
|  |  | Co | 0.17 ± 0.03 | 0.90 ± 0.03 | 0.19 |
| Bulk bcc Fe (Ref.26,32) |  | Fe | 0.09 ± 0.05 | 1.98 ± 0.05 | 0.04 |
| Bulk hcp Co (Ref.26,32) |  | Co | 0.15 ± 0.05 | 1.55 ± 0.05 | 0.10 |

Pervious work indicated that the stronger UMA for CoFeB on GaAs(100)/AlGaAs(100) is due to the enhancement of $m_{ratio}$ as they found that $m_{ratio}$ has been increased from 0.19 to 0.38 for Co and 0.34 to 0.45 for Fe when the UMA is increased from 25 Oe to 50 Oe[17,25,36]. Though the UMA in our sample was found to be as large as $270\ Oe$, the values of $m_{ratio}$ show a comparable enhancement of 0.36 for Co and 0.26 for Fe. This shows that the UMA is associated with the enhancement of the orbital moments but does not vary linearly with the $m_{ratio}$.

One of the most striking results from Table I is that the spin and orbital moments of the Co atoms are significantly larger than those of the Fe atoms. When considering the value of the $m_{ratio}$, we can see that the Co atoms also has a larger value than that of the Fe atoms. As compared the orbital moment of the crystalline hcp Co, the orbital moment of the Co atom in the CoFeB has been enhanced by more than 370%. This suggests that in the CoFeB(100) amorphous film, the Co atoms at the interface with the GaAs contribute more than Fe to the UMA. Our results indicate that the large UMA observed in the CoFeB(100)/GaAs(100) system comes from the large spin-orbit coupling of the Co atoms.

Spin-orbital coupling is a desired property in terms of the controllability by electric field in spintronic operation. The orbital moment of the Co atoms in the CoFeB/GaAs(100) has been found to be as large as 0.56 $\mu_B$, which is the largest orbital moments observed in any amorphous magnetic alloys as far as we know. It is interesting to note from Table 1 that the values of $m_{ratio}$ for the CoFeB samples on semiconductor material (SC) substrates are generally larger than that of the CoFeB samples on metal oxides (MO) such as MgO substrates. In generally, both Fe and Co values of $m_{ratio}$ for CoFeB/SC films are almost double that of the CoFeB/MO ones, although the thickness is different.

In conclusion, we have investigated uniaxial magnetic anisotropy and the elementary specific spin and orbital moments in the CoFeB(100)/GaAs system by magnetization measurement, XMCD measurement and sum rule calculations. The results obtained by VSM measurements confirmed that the UMA can achieve as large as $270\ Oe$, which is among the largest of the UMA observed in any CoFeB amorphous alloys. XMCD measurements reveal that the UMA is correlated with a strong spin-orbit coupling related to the enhanced orbital to spin moment ratios of both Fe and Co in the CoFeB. More importantly, the spin moment of the Co has been found to remain as large as that of the crystalline hcp Co, and the orbital moments is enhanced by more than 370%, suggesting the dominate contribution of the spin-orbit coupling of the Co atoms to the UMA in the CoFeB(100)/GaAs amorphous film. These results would be useful for understanding the fundamental magnetic properties of the amorphous CoFeB films, which could be important for the applications of this class of materials in the next generation spintronics devices.

*Acknowledgement*

This work is supported by the State Key Programme for Basic Research of China (Grants No. 2014CB921101), NSFC (Grants No. 61274102 and No. 61427812), NSFC if Jiangsu (Grants BK20140054), MAX-lab, Sweden and BL acknowledge support from EPSRC (EP/K03278X/I).